\documentclass[graybox,natbib,nosecnum]{svmult}
\bibpunct{(}{)}{;}{a}{}{,} 

\pdfoutput=1   

\usepackage{amsmath, amssymb}
\usepackage{mathptmx}       
\usepackage{helvet}         
\usepackage{courier}        
\usepackage{type1cm}        

\usepackage{makeidx}         
\usepackage{graphicx}        
\usepackage{multicol}        
\usepackage[bottom]{footmisc}
\usepackage[normalem]{ulem}	
\usepackage{hyperref}  
\usepackage{wasysym}
\usepackage[utf8]{inputenc}
\usepackage[T1]{fontenc}

\usepackage{soul}   

\makeindex             

\DeclareMathSymbol{\varOmega}{\mathord}{letters}{"0A}
\DeclareMathSymbol{\varSigma}{\mathord}{letters}{"06}
\DeclareMathSymbol{\varPsi}{\mathord}{letters}{"09}

\NewDocumentCommand{\mathleftmoon}{}{{\text{\normalfont\leftmoon}}}

\begin{document}


\title*{Formation of Giant Planets}
\author{Andrew N. Youdin\inst{a} and 
Zhaohuan Zhu\inst{b}}
\authorrunning{Youdin \& Zhu}
\institute{$^a$ University of Arizona, Steward Observatory and Lunar and Planetary Laboratory, Tucson, AZ, USA, \email{\texorpdfstring{youdin@arizona.edu}{}} \\
$^b$ University of Nevada, Las Vegas, NV \email{\texorpdfstring{zhaohuan.zhu@unlv.edu}{}}}
%
%
\maketitle

\abstract{Giant planets\index{giant planets} dominate the mass of many planetary systems, including the Solar System, and represent the best-characterized class of extrasolar planets\index{extrasolar planets}.  Understanding the formation of giant planets bridges the high mass end of the planet formation process and the low mass end of processes that produce stellar and brown dwarf companions.  This review examines the latest evidence supporting the formation of Solar System giant planets and most extrasolar giant planets by core accretion\index{core accretion hypothesis}.  Key elements of this theory and recent advances are discussed, along with the role of gravitational fragmentation of gas disks\index{gravitational fragmentation of disks} -- a mechanism more likely to produce brown dwarfs\index{brown dwarfs} and/or similarly massive binary companions. 
}

\section{Introduction }

This review summarizes and evaluates the current understanding of giant planet formation\index{giant planet formation}. It begins with a summary of observational constraints from the Solar System and extrasolar planets, including evidence of planets forming within disks. The leading theoretical models are then discussed, including the core accretion model and the gas gravitational instability hypothesis, more accurately described as gravitational fragmentation. Emphasis is placed on the fundamental scales relevant to planet formation, along with recent advances achieved through radiation hydrodynamic simulations. Several methods for constraining formation theories are examined, such as the initial entropy of the gas envelope, commonly referred to as 'hot starts' and 'cold starts.' The review concludes with a discussion of planetary metallicities and free-floating planets. Significant progress in theoretical and computational models of giant planet formation has been made, driven by the need to match observational breakthroughs.  Key areas for further advancement are highlighted.

\section{Observational Constraints}
\subsection{Solar System}
The Solar System's four giant planets long provided the only knowledge of planets more massive than Earth. 
The two main classes of giant planets are ``gas giants,"\index{gas giants} exemplified by Jupiter and Saturn, and ``ice giants,"\index{ice giants} exemplified by Uranus and Neptune.  Hydrogen and helium dominate the mass budget of gas giants, while heavier elements (``metals" in astronomy) make the largest contribution to ice giants.  The masses --  $317.8 M_\oplus,	95 M_\oplus,	14.5 M_\oplus$ and $17.1 M_\oplus$ for Jupiter, Saturn, Uranus and Neptune, respectively -- are frequently used as a reference for extrasolar planets (except for Uranus).  

In the context of the core accretion model, these basic features are explained by the existence of a heavy element core that accretes and retains smaller (for the ice giants) or larger (for the gas giants) amounts of nebular gas. The metallicity of Solar System giants is compared to extrasolar planets in the summary.

\runinhead{Orbits, Moons and Rings}
The orbits of the Solar Systems gas giants are quite coplanar and circular, with relative inclinations, $I \leq 1.7 ^\circ$ and eccentricities, $e \lesssim 0.05$.  These orbits suggest formation in a disk with modest dynamical excitation.  Even these modest $e, I$ rule out strong damping in planetesimal migration models \citep{morbidelli09a}.

The rotation periods of the Solar System giants are $\sim 10$-- 11 hrs for Jupiter and Saturn and $\sim 16$ -- 17 hours for Neptune and Uranus.  The faster rotation of gas giant (relative to ice giants and terrestrial planets) is plausibly due to the accretion of higher angular momentum gas from a circumplanetary disk (CPD).  Planetesimal accretion leads to slow rotation of terrestrial planets and cores \citep{DonTre93} which can be augmented by giant impacts \citep{DonTre93} and by pebble accretion \citep{takaoka23}.  (Both planetesimal and pebble accretaion are described below.)  Gas accretion from a CPD that reaches the planetary surface leads to rotation rates near breakup, i.e. matching the Keplerian speed at the equator, roughly speaking.  Jupiter and Saturn rotate at ``only" $\sim 30$ -- 40\% of breakup, implying the removal of some angular momentum, perhaps due to disk torquing by a planetary magnetic field \citep{batygin18}.

The spin axes of the giant planets have a range of obliquities, $\epsilon$, relative to the orbit plane.  Jupiter is nearly coplanar, $\epsilon = 3.1^\circ$. Saturn and Neptune have Earth-like $\epsilon = 27^\circ$ and $28^\circ$, respectively.  Uranus rotates sideways and slightly retrograde,  $\epsilon = 98^\circ$.  Dynamical spin orbit evolution can generate these obliquities \citep{harris82, lu22}.  Collisions alone are unlikely to produce these obliquities, since $e, I$ would also be excited.  However, accretion torques from late infalling gas \citep{tremaine91} or coupling with circumplanetary disks that undergo tilt instability \citep{Martin2021} could drive these obliquities, as could satellite dynamics \citep{saillenfest22, wisdom22}.  This range of possibilities precludes clear constraints on formation mechanisms.

All Solar System giant planets have rings, regular moons, and irregular moons.  Rings arise from tidally disrupted moons, and are most prominent for Saturn \citep{galileo1610}.  Regular moons are closer to their host planet with modest $e$ and $I$ (relative to the Laplace plane).  Regular moons are thought to form in a circumplanetary disk, in the later stages of giant planet formation \citep{peale15}.  Irregular moons are more distant with larger $e$ and/or $I$.  Irregular moons are thought to be captured planetesimals \citep{nesvorny07}.  

\runinhead{Meteoritic Constraints}\index{meteoritic constraints on planet formation}
Jupiter's formation history has been constrained by an interpretation of meteoritic data \citep[][hereafter K17]{Kruijer12062017}.  K17 argue that Jupiter's core reached $\sim 20 M_\oplus$ in the first 1 Myr of the Solar System, but remained  $\lesssim 50 M_\oplus$ for at least 3 -- 4 Myr.
These constraints come from (at the low mass end) the requirement to keep two different groups of meteorites -- carbon rich (CC) and carbon poor (NC) -- separated from the time of iron meteorite formation (the first $\sim 1$ Myr) until rocky, i.e.\ chondritic, meteorite formed (3 -- 4 Myr later), while also (at the high mass end) avoiding strong gravitational scattering that would mix the populations.
This explanation of the meteoritic data is only consistent with the gradual formation of Jupiter by core accretion.

However other explanations for this meteoritic dichotomy exist, including migration of the water snow line \citep{Lichtenberg2021}, the late arrival of CC pebbles from an expanding disk \citep{liu2022}, or the thermal processing of primordial pebbles due to disk outbursts \citep{Colmenares2024}.  While these other possibilities do not tightly constrain Jupiter's formation history,  they provide unique constraints on the disk's early evolution. 

\begin{figure}
\centering
\includegraphics[scale=.65]{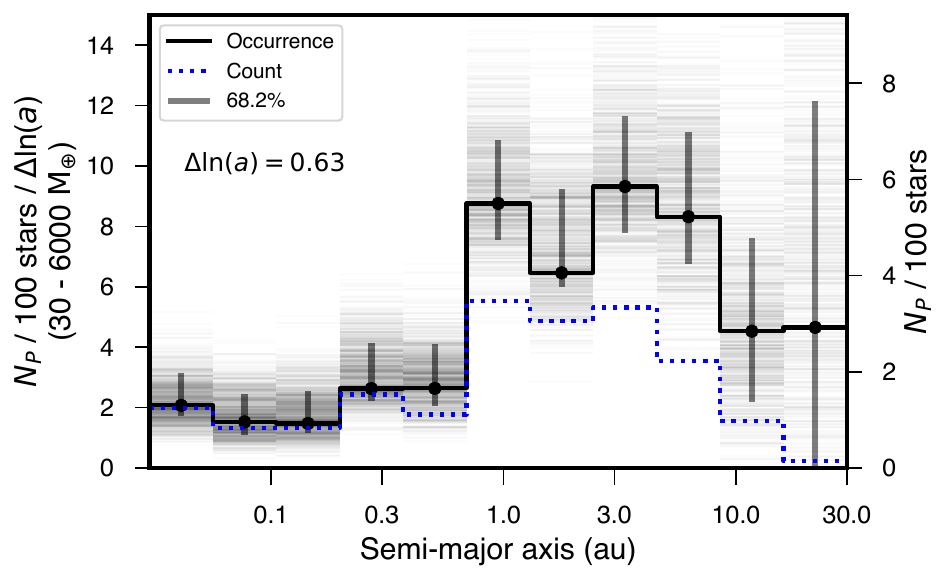}
\caption{Giant planet occurrence from radial velocity surveys, from \citet{fulton21}.  Most giant planets orbit exterior to 1 AU.  Hints that the occurrence rate drops outside $\sim 10$ AU is supported by direct imaging surveys.}
\label{fig:GP_RV}       
\end{figure}

\subsection{Extrasolar Planets}
The Solar System offers detailed characterization of a limited number of planets, while extrasolar planets provide a broader perspective on the diverse outcomes of planet formation. Key aspects of exoplanet demographics most relevant to giant planet formation are highlighted in this section. For more comprehensive overviews of exoplanet demographics\index{exoplanet demographics}, refer to \citet{zhudong21} and \citet[][this volume]{winn24}.

\runinhead{Where are the Giant Planets?}
In the Solar System, gas giants orbit at  $a \sim$ 5 -- 10 AU, and ice giants at $a \sim 20$ --30 AU.  Exoplanet surveys reveal a diversity of planetary architectures and a dependence on properties of the stellar hosts.

Radial velocity (RV)\index{radial velocity method of exoplanet detection} surveys currently have the best statistics on giant planet occurrence from short periods out to $a \sim 10$ AU.  \citet[][hereafter F21]{fulton21} present RV occurrence rates around FGKM stars for $a \leq 30$ AU, as shown in Figure \ref{fig:GP_RV}.  For giant planets with minimum masses from $\sim 0.1$ -- $20 M_{\rm Jup}$, Figure  \ref{fig:GP_RV} shows a significant increase in planetary abundance for $a \gtrsim 1$ AU.  At this transition, the occurrence rate $d N/d \ln (a)$ increases by a factor $\sim 3-4$, with $dN$ the differential number of planets per star \citep{you11b}.  With tentative $\sim 2 \sigma$ confidence, giant planet occurrence drops beyond 10 AU.  The overall giant planet abundance out to $30$ AU is $\sim 0.33$  planets per star.

F21 also confirm trends of higher giant planet occurrence around metal rich and more massive stars (F21 and references therein).  This metallicty\index{metallicity!of exoplanet host stars} bias also holds independently for giant planets around M dwarfs \citep{gan24}.  By contrast,   binary stars with separations $a \lesssim 100$ AU have the opposite trend, with fewer companions around higher metallicity primaries, at least for FGK primaries which have sufficient data \citep{moe19}.  

Direct imaging\index{direct imaging of exoplanets} surveys are sensitive to larger separation giant planets, but have many fewer detections than RV or transit surveys. The well-characterized GPIES survey  \cite[][hereafter N19]{nielsen19}, contains 6 giant planets with (modeled) masses of $5 < M/M_{\rm Jup} < 13$ and 3 brown dwarfs with  $13 < M/M_{\rm Jup} < 80$. 
While these planets all orbit more massive stars with $M > 1.5 M_\ast$, the giant planet occurrence rate for all stars surveyed 
is $\sim 3.5\pm 2 \%$ for $10 < a/{\rm AU} < 100$ (N19).
This rate is an uncertain factor of $\sim 10$ below the F21 occurrence rate at mostly shorter periods.  However since F21 included many lower mass planets (see below) and a different stellar sample, precise comparisons are difficult. 

Nevertheless, N19 independently supports the  decline in giant planet abundance with distance, by fitting $dN/d \ln (a) \propto a^\beta$ with $\beta \simeq -1.0 \pm 0.5$.  By contrast their (small number) brown dwarf sample has $\beta > 0$ (see N19 figure 8).

In summary, giant planets are most abundant between 1-10 AU and around metal rich and more massive stars.  The population distributions falls relatively sharply inside 1 AU and apparently more gradually outside 10 AU.  Stellar companions at these separations are thought to form by disk fragmentation, but have the opposite metallicity bias \citep{moe19}, suggesting that most giant planets do not form by disk fragmentation.

\runinhead{Giant Planet Masses and Sizes}

Exoplanet surveys provide constraints on the mass and size distribution of giant planets, revealing variations that depend on factors such as orbital period and stellar mass. This discussion focuses on the high-mass end, distinguishing giant planets from stellar populations, and the small-size end, which is particularly sensitive to the presence of gas envelopes.

Inside 5 AU, where RV mass distributions are more complete, F21 find a significant drop in giant planet abundances around $\sim 3 M_{\rm Jup}$ where $dN/dM \propto M^\alpha$ with $\alpha \simeq -1.7$ (obtained by fitting the high mass bins in F21 Fig.\ 6).  Similarly, L19 find a sharply declining mass distribution, with $\alpha \simeq -2.3 \pm 0.7$ for $5 < M/M_{\rm Jup} < 13$ at long periods.  
In a complementary analysis, \cite{wagner19} also find a declining mass function for directly imaged companions,  $3 < M/M_{\rm Jup} < 65$, which we roughly fit as $\alpha \simeq -1.6 \pm 0.4$.

This evidence supports a giant planet mass function that declines fairly steeply with mass above $\sim 1-10 M_{\rm jup}$, across a range of orbital distances.  This distribution differs from the rising mass function of stellar companions at the low mass end \citep{moe17}.  The giant planet mass distribution argues against a significant contribution from gas disk fragmentation, which should preferentially produce more massive companions \citep{kmy10, Zhu2012, forgan18}.

The \emph{Kepler} transit\index{transiting extrasolar planets} survey \citep{KepMission} provides the most precise measurements of planetary radius statistics. In the Solar System, the gap between the radii of ice giants ($R \sim 4 R_\oplus$) and gas giants ($R \gtrsim 9 R_\oplus$), suggests that a similar gap might be expected among extrasolar planets.
The \emph{Kepler} sample shows weak, but suggestive, evidence for a local minimum in the planet radius distribution around $R \simeq 7 R_\oplus$ for $P \lesssim 6$--8 days \citep[see their Figs.\ 3; 9, respectively]{hsu19, kunimoto20}.  At longer periods, however, these works show a declining radius distribution out to $\simeq 5$--$7 R_\oplus$, followed by a flattening.  Thus giant exoplanets are clearly less abundant than their smaller siblings, but exoplanets do not show a strong dichotomy between ice and gas giants.  This finding constrains core accretion, in general precluding a very rapid transition from small to large gas envelope masses.

At smaller sizes, \cite{lopez14} proposed $R_p \gtrsim 1.75 R_\oplus \simeq 0.5 R_\mathrm{Nep}$ as an approximate minimum radius for planets with gas envelopes $\gtrsim 1\%$ by mass,  for a range of masses $\lesssim 10 M_\oplus$.  A gap\index{radius gap in exoplanets} in the Kepler radius distribution was indeed found by \citep{fulton17} centered on $R \sim 1.7 R_\oplus$.  This gap represents a dividing between nearly bare cores and those with gas envelopes that significantly add to the radius (if not quite the mass).

The origin of this radius gap is uncertain.  It could originate at formation, dividing the planets which did and did not accrete significant gas envelopes.  However, even if most planets initially accrete significant envelopes, low mass gas envelopes can be lost due to photoevaporation, i.e.\ high energy photons \citep{owen13} and/or ``core-powered mass loss", i.e.\ heating by the cooling heavy element core \citep{ginzburg16}.  Observational surveys do not currently favor either atmospheric loss mechanism \citep{rogers21} and the initial atmosphere fraction is also difficult to constrain. 

Independent of origin, these $\gtrsim 2 R_\oplus$ "sub-Neptune" planets represent the low mass tail of giant planets, though some could have envelopes dominated by water vapor, instead of H/He \citep{zeng19}.

  \subsection{Planets in Protoplanetary Disks} \label{sec:planetsindisks}

Detecting and studying young planets in protoplanetary disks\index{protoplanetary disks} provides observational constraints on when, where, and how young planets form. However, detecting young planets using RV and transit techniques is challenging due to young stars' strong variability from gas accretion \citep{Manick2024,Donati2024}.  Since planet-disk interaction can induce disk features (e.g. gaps, spirals, velocity kinks), they have been used to constrain young planets in disks (e.g. \citealt{Zhang2018}). However, many other disk processes could also induce similar disk features \citep{bae22}. 

\begin{figure}

\centering
\includegraphics[width=2in]{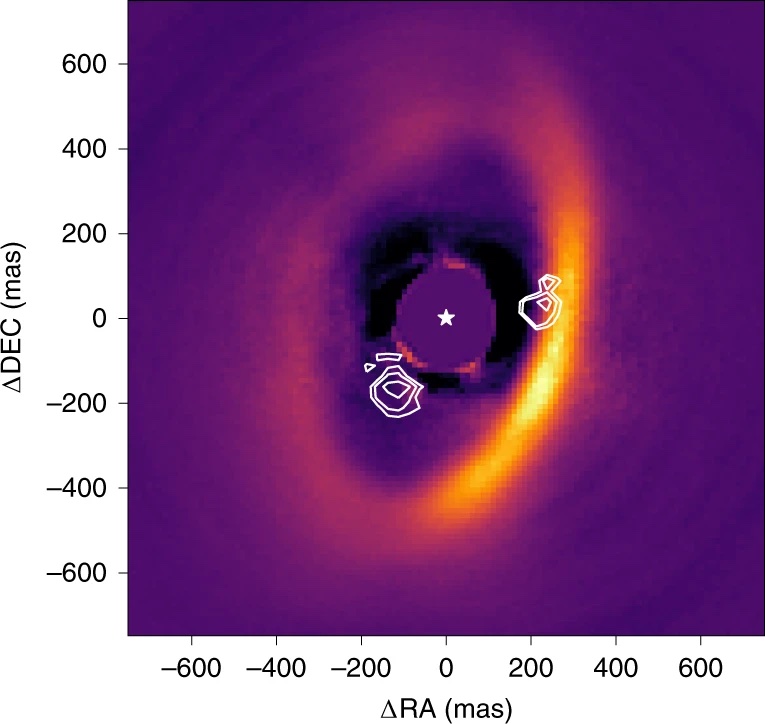}
\includegraphics[trim=0mm 0mm 67mm 0mm, clip, width=2in]{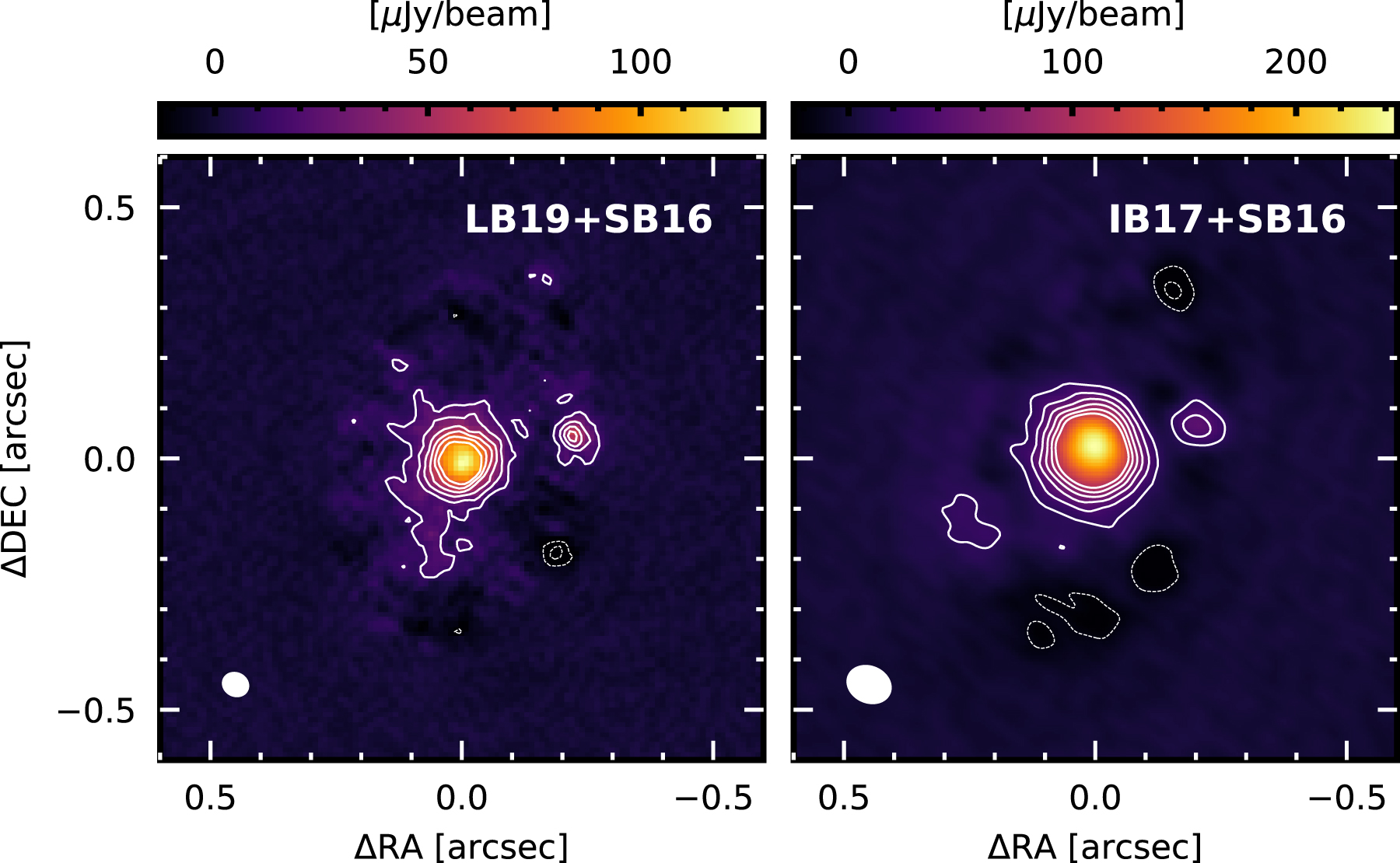}

\caption{\emph{Left:} H${\alpha}$ emission coincident with PDC 70b and c (white contours, \citealt{Haffert2019}) in the cavity of PDS 70 (SPHERE H-band image). \emph{Right:} ALMA dust continuum image of PDS70 after subtracting the outer ring, revealing the central source and an excess attributed to the PDS 70c CPD \citep{benisty21}.}
\label{fig:PDS70}       
\end{figure}

Among many planet candidates, PDS 70bc are the most robust protoplanet candidates \citep{Keppler2018,Haffert2019}. The near-IR spectrum of PDS 70b is consistent with a 1193 K and 3 $R_J$ blackbody with little molecular absorption features \citep{Stolker2020}. This radius is larger than predictions from planetary evolutionary models.  Either these models are missing important physics, or the radius is mis-estimated,  likely due to near-IR flux coming from the optically thick circumplanetary region. The H$\alpha$ emission from both PDS 70b and c suggests that each planet is accreting at $\sim 10^{-8}M_J/$yr \citep{Haffert2019}.  

ALMA detects dust continuum emission around PDS70c \citep{isella19, benisty21} which is likely from a circumplanetary disk (CPD)\index{circumplanetary disks} with a radius similar to the expected Hill radius of  PDS 70c, $\sim 0.5$--$1$ AU, consistent with tidal truncation and feeding from the circumstellar disk. The CPD dust mass appears small, 0.6--2.5 $M_{\mathleftmoon}$, depending on the assumed dust size. 
The small dust mass and very small gas accretion rate --  corresponding to modest dust delivery rate $\sim 3 \times 10^{-3}M_\mathleftmoon/{\rm Myr}$ for a 1\% dust mass fraction -- suggests that any significant moons formed already.  For a higher rate of dust supply, low mass CPDs could form moons \citep{CanupWard2002}.
Finally, planet-disk interaction theory can successfully reproduce many observed features in PDS 70 system, including gas and cavity sizes, possibly resonant planets, gas accretion, etc. \citep{Bae2019}.
  
Accretion disks around free-floating \index{free-floating planets} and wide-orbit planetary mass companions may shed light on giant planet formation, while also representing a miniature version of star formation processes. For example, SR 12c is a $\sim 11$--$14 M_{\rm Jup}$ companion at 980AU from the T-Tauri binary SR 12AB. SR 12c hosts a dust emitting disk with radius $\lesssim 5$  AU \citep{wu22}.  This disk is thus much smaller than the Hill radius, $R_{\rm H} \sim 150$ AU of SR 12c, while its mass is close to the CPD around PDS 70c. On the other hand, the CPD mass around free-floating OTS44 ($\sim 11.5 M_{\rm jup}$) is $\sim$5 times higher \citep{Bayo2017}.
For context, the more massive OTS44 disk has an estimated dust-to-central object mass ratio that is $\lesssim 0.1$ the Solar MMSN value, comparable to late stage class II/III circumstellar disks \citep{andrews13}.

\section{Theoretical models}
\subsection{Scales of the problem}
To understand how giant planets form around stars of different mass $M_\ast$ at different radial locations, $r$, in disks, it is instructive to consider the fundamental  mass, length and time scales of relevance \citep{raf06, youdin13}.

Orbits in disks have a nearly Keplerian orbital period, $P$, and an orbital frequency $\varOmega = 2 \pi /P = \sqrt{G M_\ast/r^3}$, with $G$ the gravitational constant.  Disk gas is mostly located within a pressure scaleheight $H$ of the disk midplane, with $h = H/r$ the aspect ratio.  The sound speed is $c_{\rm s} \simeq H \varOmega$, due to vertical hydrostatic balance.  In a smooth disk, pressure has an $\mathcal{O}(h^2)$ effect on orbital speeds, while disk self gravity has a $\mathcal{O}(M_{\rm disk}/M_\ast) \sim h/Q$ effect, for Toomre's $Q$ (below).  

A planet of mass $M_{\rm pl}$ can bind disk gas within a Bondi radius\index{Bondi radius} 
\begin{align}
 R_{\rm B} = GM_{\rm pl}/c_{\rm s}^2, 
\end{align} 
where the planet's gravitational potential energy exceeds the gas thermal energy.  Disk gas will also remain unbound outside the Hill radius\index{Hill radius}
\begin{align}
 R_{\rm H} = \left(\frac{M_\mathrm{pl}}{3 M_\ast} \right) ^{1/3} a, 
\end{align} 
due to the star's tidal gravity.

These length scales (nearly) match, $R_{\rm B} \simeq R_{\rm H} \simeq H$, when  the planetary mass, $M_{\rm pl}$, equals the thermal mass,\index{thermal mass} 
  \begin{align}
M_{\rm th} &= c_{\rm s}^3/(G \varOmega) = h^3 M_\ast \label{eq:Mth_h}\\
&\simeq 25  \left(\frac{r}{10~\mathrm{AU}} \right)^{6/7} M_\oplus. \label{eq:MthN} 
\end{align}
using an irradiated disk model with $M_\ast = M_\odot$ for the numerical estimate of $c_{\rm s}$ \citep{cg97, cy10}.
Subthermal planets, $M_{\rm pl} \lesssim M_{\rm th}$, mostly accrete from the disk midplane, since $R_{\rm B} \lesssim R_{\rm H} \lesssim H$.  Superthermal planets, $M_{\rm pl} \gg M_{\rm th}$, see more of the disks vertical stratification, since $H \lesssim  R_{\rm H} \lesssim R_{\rm B}$.

A solid core of density $\rho_{\rm c}$ and radius $R_{\rm c}$ can bind a gas envelope if $R_{\rm B} > R_{\rm c}$, i.e.\ if $M_{\rm pl}$ exceeds
\begin{align}
M_{\rm B} &= \frac{c_{\rm s}^3}{\sqrt{4 \pi G^3 \rho_{\rm c}/3}}  
= \frac{\varOmega}{\sqrt{4\pi G \rho_{\rm c}/3}}M_{\rm th}\\
&\simeq 5 \times
10^{-6} \sqrt{\frac{M_\ast}{M_\odot}} \left(\frac{r}{10~{\rm AU}}\right)^{-3/2} M_{\rm th}
\end{align} 
for $\rho_{\rm c} = 6~{\rm g/cm^3}$.  Planets can start gas accretion well below the thermal mass.    

For self-gravitating disks with surface mass density, $\varSigma_{\rm g}$, the relevant lengthscales are due to pressure $\lambda_P = c_{\rm s}^2/ (\pi G \varSigma_{\rm g})$ and angular momentum $\lambda_J = \pi G \varSigma_{\rm g} / \varOmega^2$.  Thin disks are axisymmetrically gravitationally unstable when $\lambda _P < \lambda_J$, i.e.\ when Toomre's\index{Toomre's $Q$} $Q = c_{\rm s} \varOmega/(\pi G \varSigma_{\rm g}) < 1$.  At marginal stability $Q \simeq 1$, and $\lambda_P \simeq \lambda_J \simeq H$.  

When disks cool rapidly they produce fragments with a dimensionally estimated mass, $\pi \varSigma_{\rm g} H^2 = M_{\rm th}/Q \simeq M_{\rm th}$,  i.e.\ the thermal mass, with an uncertain prefactor.    Simulations give the average value of this prefactor, so the initial fragment mass is  $\sim 45 M_{\rm th}$ \citep[see below]{xu24}.  Thus in a warm gravitationally unstable disk with  $h \simeq 0.1$,  Eq.\ (\ref{eq:Mth_h}) gives initial fragment masses $\sim 45 M_{\rm Jup}$.  Disk fragmentation is discussed further below. 
 
\begin{figure}
    \centering
    \includegraphics[width=4.65in]{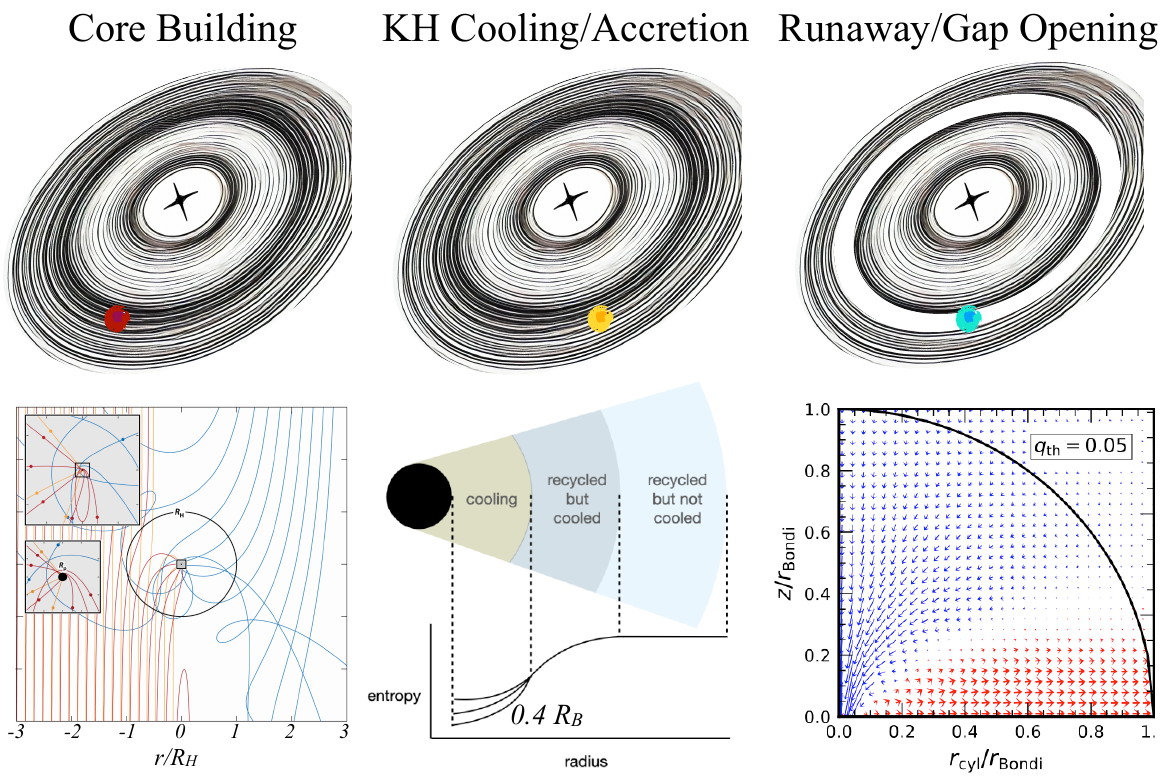}
    \caption{Three stages of giant planet formation by core accretion. (\emph{Left:})  The heavy element core grows, accreting planetesimals and pebbles onto a seed planetesimal.
    The lower left panel \citep{johansen17} shows trajectories for  different particle sizes and Stokes numbers (blue:  
    $\tau_{\rm s} \gg 1$, red, orange: $\tau_{\rm s}$=1.0, 0.1) in the shear-dominated accretion regime. 
    (\emph{Middle:}) The gas envelope accretes via cooling while attached to the disk.  The lower middle panel shows the effect of recycling flows on the structure of the accreting gas envelope  
    \citep{Bailey2024}. 
    (\emph{Right:})  The envelope detaches from the disk during runaway growth and contraction. Accretion slows as the planet opens a gap.\index{gaps in disks}  The lower right panel  \citep{Choksi2023} shows Bondi-like accretion onto the poles and outflow in the disk plane.  The presence or (in this case) absence of central sinks to allow accretion affects the outflow.   
    (\emph{Observability:})  Before runaway growth, the embedded planet locally heats and brightens the disk \citep{Zhu2023}. After gap opening, the forming planet and its CPD may be detectable with less reddening.  See text.}
    \label{fig:evolution}       
\end{figure}

\subsection{Growth of Heavy Element Cores}
In the core accretion hypothesis\index{core accretion hypothesis} (Figure \ref{fig:evolution}), rocks and ices in the protoplanetary disk first grow a heavy-element core.  The core must grow within the  $\sim 1$--10 Myr median lifetime of protoplanetary disks \citep{pfalzner22}, on average, to allow time to accrete disk gas.

\runinhead{Planetesimal Formation: over the meter-size barrier}\index{planetesimal formation}
In the core accretion hypothesis, the journey to become a giant planet begins with collisions between small dust grains. 
The complex story of how 1--100 km solid planetesimals form is summarized below (also see reviews by \citealp{cy10, johansen14, klahr18HE, lesur23}).  For giant planet formation, relevant outcomes include both the properties of the planetesimal building blocks, and also the remnant fraction of dust and pebbles -- important sources of opacity and mass for subsequent growth.

The growth of dust grains by sticking slows -- perhaps to insignificant rates -- at $\sim$ mm or cm sizes (of compacted grains), due to bouncing, erosion and/or fragmentation \citep{birnstiel16}.  Furthermore
if solids grow to $\sim$ 10 cm -- m sizes, they would radially drift\index{radial drift of dust} into the star in $\sim 100$ years due to pressure supported gas \emph{in a smooth disk} \citep{ahn76, houches10}.

To overcome barriers to growth beyond $\sim$ meter-sizes, a cloud of pebbles could gravitationally collapse directly into large planetesimals \citep{saf69, gw73, ys02}.
Several mechanisms, which may act in concert, could increase pebble densities and trigger collapse.  First, pebbles collect in pressure bumps -- either rings or vortices.  These substructures have apparently been observed (reviewed in \citealp{pinilla17, bae22}).  

Also, the streaming instability (SI)\index{streaming instability} causes drifting pebbles to spontaneously clump due to the feedback of the particle drag forces on gas \citep{yg05, jy07}. The SI has been extensively simulated (see above reviews) and shown to trigger the formation of $\sim 10-1000$ km planetestimals by gravitational collapse.    In the related secular GI mechanism, drag forces mediate a slow gravitational collapse of pebbles into rings, giving an alternate explanation of disk substructures \citep{you11a, tominaga20}.

This discussion suggests that the understanding of planetesimal formation is influenced by the interpretation of disk substructures.  This interpretation is complicated by the fact that already-formed planets may be the leading cause of observed features (see observational overview, \citealp{Zhang2018}).  High resolution velocities of disk gas could test the pressure bump hypothesis (due to disk gaps or other hydrodynamics, see \citealp{bae22}), with the observable properties of gas rings subject to a Rossby wave stability constraint \citep{chang23}.

In summary, the leading hypothesis for planetesimal formation is a multi-step process where dust grains first grow to larger pebbles. Pebbles are collected -- by the SI and/or pressure bumps -- to densities that trigger gravitational collapse into planetesimals as large as $\sim 1000$ km.  A significant dependence on uncertain disk conditions exists.  Observational support for this physically complex picture exists.  Pebbble cloud collapse -- as seen in SI simulations \citep{li19} -- reproduces the observed matching colors and the primarily prograde orbits of primordial Kuiper Belt \index{Kuiper Belt Objects} binaries \citep{nyr10, nesvorny19}.

\runinhead{Growing Cores: General Considerations}

Before accreting a massive gas envelope, a non-volatile core must first grow to $\sim 10 M_\oplus$, a value that depends on various factors (see below).  Cores which fail to accrete, or subsequently lose, significant gas envelopes wind up as rocky planets.

Starting with $\sim 10$--1000 km planetesimals, core growth proceeds by accreting solids of various sizes.  For the accretion of small pebbles ($\lesssim 1$ m), gas drag is important during the encounter with the core \citep{ormel10}.  Larger planetesimals have gravity-dominated encounters, but gas can damp their velocity dispersion and encounter speed \citep{rafikov04}.

The basic mass accretion rate of a protoplanet (labelled $i$) is given by
\begin{align}\label{eq:mdotcore}
\dot{M}_i &= \sum_j \rho_{j} \sigma_{ij} \Delta v_{ij} 
\end{align} 
for the accretion of all possible bodies $j$, with mass densities $\rho_j$.  The accretion  cross section, $\sigma_{ij}$, is larger for massive bodies due to gravitational focusing.  The random or encounter speed $\Delta v_{ij}$ gives the mass flux $\rho_j \Delta v_{ij}$.

Smaller speeds,  $\Delta v_{ij}$, also increase gravitational focusing and $\sigma_{ij}$, typically leading to a larger overall $\dot{M}_i$ \citep{gls04}.  Colloquially, ``cool food can be eaten quickly."  Contributions to $\Delta v_{ij}$ include random velocities stirred by gravitational encounters, the aerodynamic drift of pebbles and -- at the low end for the fastest ``shear-dominated" accretion -- the Keplerian shear of the disk.

Some generalizations to Eq.\ \ref{eq:mdotcore} are needed in practice.  Collisions resulting in cratering or disruption require loss terms \citep{leinhardt12}.  Dynamics becomes two-dimensional in a thin disk, where the bodies have scaleheights $H_i^2 + H_j^2 < \sigma_{ij}$.   For 2D accretion, the relevant mass per mean free path becomes, $\sigma_{ij} \rho_j \rightarrow 2 b_{ij} \varSigma_{j}$ using the surface density $\varSigma_{j}$ and collisional impact parameter $b_{ij} = \sqrt{\sigma_{ij}/\pi}$.  

Speeds and/or collision probabilities, $p$ that depend on impact parameter (as for shear-dominated accretion) are included as $\Delta v_{ij} \sigma_{ij} \rightarrow \pi \int p(b) \Delta v_{ij}(b) b db$ or, for 2D accretion, $2 \Delta v_{ij} b_{ij} \rightarrow  \int p(b) \Delta v_{ij}(b) db$, which reduces to the simple case when $p(b)$ is an amplitude 1 step function out to $b = b_{ij}$.

\runinhead{Planetesimal Accretion onto Cores}\index{planetesimal accretion}
At large orbital distances, the growth of  $\sim 10 M_\oplus$ cores by planetesimal accretion is limited by the accretion timescale, $t_{\dot{M}} = M_{\rm pl}/\dot{M}_{\rm pl}$.   Accretion is fastest in the shear dominated regime, when $\Delta v_{ij} \lesssim \varOmega R_{H,i}$.  In this case,  $t_{\dot{M}} \lesssim 1$ Myr out to $\sim 25$ AU in a minimum mass nebula and and out to $\sim 70$ AU with a 5 times increase in solids \citep{youdin13}.  These rates are an upper limit for small planetesimals with strongly collisionally damped random speeds \citep{gls04}.  Slower rates that restrict core growth to shorter separations apply to larger, less damped planetesimals.  At small orbital distances, the more relevant limit is isolation mass,\index{isolation mass!to planetesimals} $M_{\rm iso}$, where \emph{planetesimal} accretion stalls once all planetesimals within several $R_\mathrm{H}$ are consumed.  For the fastest accretion, the feeding zone is only $\sim 2 R_\mathrm{H}$ wide and isolation masses are quite small \citep{rafikov04}.  However with larger velocity dispersions, the feeding zone spans $\sim 10 R_{\rm H}$ \citep{ki98}.  Then  $M_{\rm iso} \gtrsim 10 M_\oplus$ for $r \gtrsim 4$ AU in a minimum mass nebula disk or for $r \gtrsim 0.6$ AU for a factor 2.5 increase in solids (not 5 as above, accounting for loss of ice inside the snowline).

In summary, core formation by planetesimal accretion is possible over a range of intermediate disk radii, which corresponds the orbital locations of most observed giant planets.  However, consistent calculations of planetesimal stirring and damping are needed for realistic accretion rates. 

\runinhead{Pebble Accretion onto Cores}\index{pebble accretion}
Aerodynamic pebble accretion can give faster core growth, which is not limited by a planetesimal feeding zone \citep[see][this volume]{ormel24}.

Due to the drift of pebbles from the outer disk \citep{lambrechts14}, the planetesimal isolation mass (above) does not apply.  Instead a growing core isolates itself from pebbles by carving a gap,\index{gaps in disks} with a pebble-trapping pressure bump on the outer edge \citep{MorNes12}.  This pebble isolation mass\index{isolation mass!to pebbles} depends on many disk and planet parameters, including the effective viscosity parameter $\alpha$ \citep{armitage_2020},
but a rough estimate is the gap-opening mass \citep{Zhu2013}
\begin{equation}
\frac{M_p}{M_*}\sim \sqrt{6\alpha}\left(\frac{H}{r_p}\right)^{5/2}\sim 14 M_{\oplus}\left(\frac{\alpha}{10^{-3}}\right)^{1/2}\left(\frac{h}{0.05}\right)^{5/2}\, .
\end{equation}
While very low-mass planets could open gaps in an inviscid disk, the opening timescale becomes too long.  \cite{Bitsch2018} gives detailed fitting formulas for the pebble isolation mass in viscous disks, accounting for the turbulent diffusion of pebbles in the trap.

If the pebble isolation mass is large enough, significant envelope accretion could occur despite heating by pebble accretion.  Isolation could be imperfect, since trapped pebbles should form planetesimals and planets \citep{hu18, jiang23}, which might get accreted. 

Pebble accretion can be  faster than planetesimal accretion.  Comparing the fastest shear-dominated limit (using Eq.\ 7.13 and 7.15 of \citealp{ormel17}), the ratio of pebble to planetesimal accretion rates is
\begin{align}
\frac{\dot{M}_{\rm peb}}{\dot{M}_{\rm ptml}}  &\approx \frac{2 \tau_{\rm s}^{2/3}}{11} \sqrt{\frac{R_{\rm H}}{R_{\rm p}} }  \frac{\varSigma_{\rm peb}}{\varSigma_{\rm ptml}}  \approx 8 \tau_{\rm s}^{2/3} \sqrt{\frac{r }{10~{\rm AU}}} \frac{\varSigma_{\rm peb}}{\varSigma_{\rm ptml}}
\end{align} 
which depends on the surface density ratio of pebbles to planetesimals ${\varSigma_{\rm peb}}/{\varSigma_{\rm ptml}}$ and the Stokes number, $\tau_{\rm s} = \varOmega t_{\rm s}$, a scaled stopping time $t_{\rm s}$ \citep{houches10}. This estimate, and pebble accretion, is valid for tightly coupled pebbles with $\tau_{\rm s}\lesssim 1$.
The final numerical scaling drops a weak $[\rho_{\rm c}M_\odot/(M_\ast \cdot 6{\rm g\; cm^{-3}}\;)]^{1/6}$ dependence. 

From this comparison of only the fastest rates, it appears that pebble accretion is slower at small $\tau_{\rm s}, r$ or ${\varSigma_{\rm peb}}/{\varSigma_{\rm ptml}}$.  However, the reason pebble accretion is thought to be much faster is the difficultly of damping planetesimal velocities to shear-dominated values.  This caveat emphasizes the need for detailed modeling of planetesimal and pebble accretion \citep[e.g.\ ][]{liu19}. 

While pebble accretion is fast, especially as $\tau_{\rm s} \rightarrow 1$, it also becomes inefficient in the accreted fraction of the pebble flux
\begin{align}
\epsilon & = \frac{\dot{M}_{\rm peb}}{\dot{M}_{\rm drift}}  = \frac{R_{\rm H}^2 \varOmega}{2 \pi r v_{\rm hw} }  \tau_{\rm s}^{-1/3} \\
&\simeq \frac{0.014}{\tau_{\rm s}^{1/3}} \left(\frac{M_{\rm p}}{10~M_\oplus} \right)^{2/3} \left(\frac{M_\odot}{M_\ast} \right)^{1/6} \left(\frac{50~{\rm m/s}}{v_{\rm hw}} \right) \sqrt{\frac{10~{\rm AU}}{r} } \nonumber
\end{align} 
again for shear-dominated accretion with a typical headwind speed $v_{\rm hw}$ from radial pressure gradients.  This inefficiency allows the disk's pebble flux to fuel the growth of multiple cores.  

The gas envelopes bound to growing cores further facilitate the capture of small planetesimals \citep{InaIko03} and pebbles \citep{dangelo24}.  The most rapid growth of cores combines planetesimal and pebble accretion with the effects of a gas envelope.

\begin{figure}
\centering
\includegraphics[scale=.4]{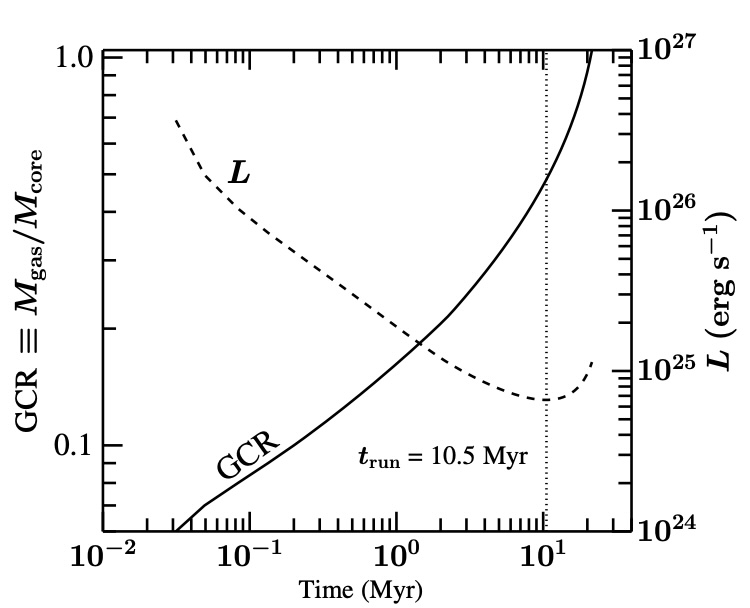}
\caption{The gas envelope to core mass ratio (GCR, solid curve) and luminosity (dashed curve) for a $5 M_\oplus$ core at 0.1 AU \citep[from][]{lee14}. The cooling luminosity drops until GCR reaches 0.5.  Subsequently, the gas accretion accelerates into runaway growth.}
\label{fig:GCR_L}       
\end{figure}
        
\subsection{Accretion of Gas Envelopes}
 In the core accretion hypothesis\index{core accretion hypothesis}, heavy element cores accrete gas from the protoplanetary disk. 
The general processes of accretion onto objects orbiting in gas disks applies more broadly, even to other astrophysical disks.

\runinhead{Gas Accretion in 1D}
Standard core accretion models represent the protoplanet as a one-dimensional, spherically symmetric body \citep{BodPol86}. The evolution of the gas envelope is described using the stellar structure equations, which do not account for hydrodynamic disk flows. Gas accretion proceeds in two distinct stages (Figure \ref{fig:evolution}). During the initial attached phase, the gas envelope transitions smoothly to the disk, typically matching midplane conditions at
${\rm min}(R_{\rm B}, R_{\rm H})$ \citep{rafikov04}.   In the subsequent detached phase, gas flows dynamically onto the planet, passing through a surface accretion shock and/or a CPD. These complex flows must be spherically averaged to be incorporated into one-dimensional models. The two phases and the transition between them are described below.

During the attached phase, the envelope growth is regulated by the envelope's ability to cool, i.e.\ growth occurs on a Kelvin-Helmholtz timescale\index{Kelvin-Helmholtz timescale} \citep{Piso2014}.   ``To cool is to accrete'' \citep{lee15} means that the envelope mass, $M_{\rm env}$ grows at a rate $\dot{M}_{\rm env} \simeq M_{\rm env}/t_{\rm cool} \simeq M_{\rm env} L_{\rm cool}/U_{\rm env}$ where $U_{\rm env}$ is the internal energy of the envelope. The envelope's cooling luminosity, $L_{\rm cool} = L - L_{\rm c} - L_{\rm heat}$, is the total planetary luminosity, $L$, minus contributions from the core (radioactivity or cooling), $L_{\rm c}$, and from heating, $L_{\rm heat}$, primarily due to solid accretion.   Thus heating by planetesimal and pebble accretion inhibits envelope cooling.
The large solid accretion rates needed to grow the core generally need to taper off before significant envelope accretion occurs \citep{pollack96}.

Cooling times -- and thus $\dot{M}_{\rm env}$ -- also depend on the core mass, envelope opacity $\kappa$, and disk temperature \citep{ikoma00, piso15}.  Higher core masses and lower opacities produce more rapid envelope growth.  These trends are simply explained by $L_{\rm cool} \propto M_{\rm pl}/\kappa$ for radiative cooling from the envelope's convective zone (which initially dominates cooling, \citealp[]{Piso2014}).  
Trends with disk temperature are more complicated, as disk temperatures affect both radiative losses and the envelope mass \citep{lee15}.  Overall, lower temperatures increase $R_{\rm B}$, accelerating the growth of massive envelopes \citep{Piso2014}.

Gas accretion by cooling is a ``bottleneck" process, where the rate limiting step occurs just before the envelope becomes self-gravitating.  Figure \ref{fig:GCR_L} shows this effect with a cooling luminosity ($L = L_{\rm cool}$ here) that decreases with mass for low mass envelopes, before turning over and increasing with mass for $M_{\rm env} \gtrsim 0.5 M_{\rm core}$.  This luminosity minimum corresponds to the onset of runaway growth\index{runaway growth of giant planet envelopes}.  Envelope growth becomes exponential in time not quite at the luminosity minimum, but at slightly higher masses when $t_{\rm cool} = U_{\rm env}/L_{\rm cool}$ reaches a maximum, after which growth is slightly super-exponential.

The initially decreasing $L_{\rm cool}(M_{\rm env})$
occurs as the radiative zone extends to deeper pressures, an opacity-dependent effect also known from hot Jupiter evolution \citep[][]{ym10}. This trend reverses, and $L_{\rm cool}(M_{\rm env})$ increases, due to the increased cooling from self-gravitating envelopes, especially the growing outer radiative zones \citep{Piso2014}.

A crucial issue is the minimum core mass needed for runaway envelope growth to occur within the gas disk lifetime.  This ``critical core mass" (often defined in slightly different ways) will decrease for any effect that increases $L_{\rm cool}$, notably lower solid accretion rates and envelope opacities.  While values of $\sim 5-20 M_\oplus$ are ``standard," many works evaluate $M_{\rm crit}$ for different parameters \citep[e.g.][]{ikoma00, piso15}. 

Runaway (super) exponential mass growth cannot persist for long.  The attached phase ends when the required gas accretion rate exceeds what the disk can supply.  The maximum supply rate is limited by thermal Bondi accretion\index{Bondi accretion} \citep{ginzburg19, Choksi2023}, by the disks (effective) viscous accretion rate  \citep{HubBod05} and the opening of deep gaps\index{gaps in disks} in the disk by the growing planet  \citep{LisHub09, ginzburg19}.  While 1D models set this limiting luminosity by hand, 3D simulations try to measure a consistent value.
    
\runinhead{Gas Accretion in 3D, Circumplanetary Disks}\label{sec:3D}index{circumplanetary disks}
While 3D simulations are useful in determining the accretion rate onto detached planets after runaway growth, they are also important for the attached phase (see Figure \ref{fig:evolution}).  Complex flows that exchange material between the gas envelope and disk (across the Bondi and Hill radii) are neglected in 1D and require 3D simulations  of planet-disk interactions to accurately model.   

These 3D numerical simulations show that the flow pattern around embedded planets is highly complex, potentially affecting envelope accretion through Kelvin-Helmholtz contraction. Isothermal simulations have shown that gas in the disk flows to the planet from the pole, and then leaves the planet from the midplane \citep{Machida2008, Tanigawa2012, Fung2015, Ormel2015,Bethune2019}.  As discussed above, envelopes attached to the disk must cool to accrete.  And the timescale to lose envelope entropy is the bottleneck to accreting a massive envelope.
The recycling in 3-D simulations mixes some of the cooling, lower entropy envelope material with higher entropy disk material, lengthening cooling times and potentially preventing run-away accretion.

The impact of the 3-D flows depends on how deeply they extend into the planetary envelope. While some works suggest that significant recycling extends all the way to the planetary core \citep{Moldenhauer2021}, others find that there may be a bound inner envelope that is not affected by recycling  \citep{Lambrechts2017}.  Radiation hydrodynamic models in \citet{Bailey2024} find that only the outermost layers of the envelope are strongly prohibited from cooling.  An intermediate layer can cool modestly despite recycling, while the inner $\sim 0.4 R_{\rm B}$ can cool as in 1D models (see Fig.\ \ref{fig:evolution}). This structure would somewhat delay runaway envelope growth, especially for the {\it in situ} growth of hot Jupiters, with smaller $R_B$.  In general, 3D recycling should  slightly reduce the parameter space for runaway growth.\index{runaway growth of giant planet envelopes}

In both the attached and post-runaway phases, 3-D simulations show that the flow structure sensitively depends on the treatment of thermodynamics. Radiation hydrodynamical simulations \citep{Ayliffe2009,DAngelo2013,Szulagyi2016,Cimerman2017,Lambrechts2017} generally reveal that the circumplanetary region becomes a pressure supported atmosphere instead of a rotationally supported disk in isothermal simulations (above). 
From radiation hydrodynamic simulations that include spatially-varying dust opacities, \citet{krapp24} infer a cooling time criterion for the formation of rotationally supported CPD structures. Future 3-D simulations should further incorporate dust evolution and dynamics (including pebble isolation) in the radiation hydrodynamical simulations to self-consistently model planet growth and CPD accretion, since radiative cooling depends on dust size and dust abundance.

\subsection{Gravitational Instability and Fragmentation of Gas Disks} \label{sec:GI}\index{gravitational instability hypothesis|see{gravitational fragmentation of disks}}

The gravitational fragmentation\index{gravitational fragmentation of disks} of a circumstellar disk can produce stellar companions and is a leading mechanism for forming stellar binaries at separations of 
 $\lesssim 100$ AU \citep{offner23}. Whether the low-mass end of disk fragments extends to giant planets remains a longstanding question. The emerging observational constraints summarized above suggest that this process is rare. The following discussion focuses on the theoretical understanding of this mechanism.

Young massive disks, fed by infalling gas from a molecular cloud core, become gravitationally unstable when $Q \lesssim 2$, i.e.\ for disk masses $M_{\rm d} \simeq \pi \varSigma r^2 \gtrsim h M_\ast/2$ \citep{sellwood84, kratter2010}. 
Gravitational instability generates ``gravito-turbulence" and or larger scale spirals, which drive accretion and  heat the disk.  Unstable disks may also fragment into bound companions.  These responses regulate the disk mass and temperature, raising Toomre's $Q$\index{Toomre's $Q$} back to marginal stability \cite[as reviewed by][]{kratter16}.  

Disk fragmentation is most likely when the cooling time is sufficiently short, $t_{\rm cool} \lesssim 3/\Omega$, in either 2D or vertically stratified disks \citep{gam01, baehr17}.   High resolution simulations reveal fragmentation as a stochastic process that occurs, with strongly decreasing probability, at longer cooling times \citep{paardekooper12, brucy21}.  The (fuzzy) cooling time requirement for fragmentation means that disks should fragment beyond $\sim 50$ AU (depending on disk properties), where they become at least marginally optically thin \citep{raf05, clarke09, kmy10}.

Rapid cooling in marginally gravitationally unstable disks does not guarantee fragmentation. 
To reach a violently unstable disk state, with $Q$ values low enough for fragmentation, requires high rates of mass infall and low rates of angular momentum infall \citep{kratter2010, kratter16}.  This dependence on infall occurs even for isothermal simulations with zero cooling time \citep{kratter2010} and also for simulations with more realistic opacities \citep{offner10, Zhu2012}. Many simulations neglect infall, but instead start in a violently unstable, low $Q$, state.  Such simulations are informative, but do not address how, i.e.\ at what infall rates, disks get to those states.

If and when fragmentation occurs, the initial masses of fragments is difficult to determine due to the computational challenge of modeling 3D self-gravitating disks with realistic radiative cooling and infall \citep{krumholz07}.  Radiation hydrodynamic models appear to produce different results, and it is difficult to understand which differences are due to physics vs.\ numerics.  For example \citet{steiman-cameron23} find enhanced transport but no fragmentation.  However, \citep{boss21} produce fragments at surprisingly short orbital distances, using the opacities of \citep{boss84}.  These opacities likely misrepresent realistic disk conditions since the mean $\kappa_R \simeq 0.01 ~{\rm cm^2/g}$ at $T \simeq 200$ K (shown in Fig.\ 1 of \citealp{boss84}) is a factor $\sim 300$ below more standard values in \citet{SemHen03} (c.f.\ their Fig.\ 1).

The numerical experiments of \cite{xu24}  provide a useful guide to fragment masses by considering constant (grey and temperature independent) opacities for both optically thin and marginally thick ($\tau = 10$ at the outer edge) annuli, using the discrete ordinate method of \citep{jiang21}.  \cite{xu24} find critical cooling times (beyond which fragment production is rare) of $\simeq 5.5/\Omega$ and $2/\Omega$ for the optically thin and thick cases, respectively.  The average fragment mass was $\simeq 45 M_{\rm th}$,\footnote{Obtained by applying the mean $M_{\rm frag} \simeq 40 h^3(M_\ast + M_{\rm disk})$ and $M_{\rm disk} = 0.1 M_\ast$ in \cite{xu24}.} as noted in the above introduction to fragment masses.  \cite{xu24} fit a log normal mass distribution, which produces fragments $< 1/3$ the mean mass only 3\% of the time.  Averaging over parameters for young disks \citep{xu22} yields a mean physical fragment mass of $20~ M_{\rm jup}$.

Fully realistic simulations are challenging, but would include include temperature dependent opacities, disk irradiation (as in the shearing box study of \citealp{hirose19}) and infall onto the disk.
Crucially, the initial fragment masses in brief simulations (usually a few orbits) is unlikely to represent the final companion mass accurately.  In a massive disk with ongoing infall, significant clump evolution will occur \citep{Zhu2012}.  This evolution is generally unfavorable for the survival of giant planets, including: significant gas accretion to grow to a stellar binary companion \citep{kmy10}, rapid stochastic migration \citep{Baruteau2011},  which can lead to tidal destruction in the inner disk \citep{Boley2010}.

Alternately, tidal disruption may be incomplete and result in a surviving ``downsized" planet \citep{Nayakshin2010}.  This possibility is quite difficult to simulate consistently, and reviewed in \citealp{kratter16}.  Population synthesis models\index{population synthesis models} suggest that significant downsizing is rare, and that most downsized objects are ejected \cite[see their figure 3]{forgan18}.

Combining these theoretical constraints with the observational constraints discussed above, it seems unlikely many giant planets $\lesssim 10 M_{\rm jup}$ form by the gravitational fragmentation of gas disks\index{gravitational fragmentation of disks}.

\subsection{Initial Entropy of Giant Planets}\index{entropy of giant planets}
The  entropy of a giant planet is a crucial parameter, since its luminosity is mainly generated by entropy loss, or cooling.  (Radioactive decay and gravitational settling of heavy elements play a smaller role.) 
Since initial entropy is set during formation, observed luminosities might constrain formation. 

High initial entropy or ``hot starts"\index{hot starts} have Kelvin-Helmholtz cooling times comparable to the formation timescale of $\sim$ Myr.  Low initial entropies or  ``cold starts"\index{cold starts}, assume a perfectly efficient accretion shock, as in \cite{bodenheimer00}, see below.  The dividing line between these hot and cold starts is $\sim 9.5 k_B/$baryon.  Hot and cold starts lie further from this mean with increasing mass, as a ``tuning fork" \citep{marley07}.

Disk fragmentation is assumed to produce more luminous hot starts (though this outcome is challenging to simulate in detail).   While earlier core accretion models produced the definition of cold starts \citep{bodenheimer00}, more recent studies show that core accretion models produce warm or even hot starts \citep{mordasini13,berardo17, marleau19, chen22}.   While initial entropy is not a clear indicator formation mechanism, it does depend on accretion rates and other formation details.

Ignoring angular momentum (for now, see below), massive giant planets likely accrete most of their mass through an accretion shock.  The entropy delivered through this radiating shock would then determine the starting entropy.   Often the full accretion luminosity, $L_{\rm acc} = GM_{\rm pl} \dot{M}/R_{\rm pl}$, is assumed to radiate away \citep{bodenheimer00}, with none of this energy directly delivered into the planet's envelope.  In this case of a perfectly efficient shock, the accreted entropy has a minimum value, set by the density and temperature at the base of the shock, which also serves as the planet's photosphere.

\cite{mordasini13} showed that even for standard efficient shocks, higher gas accretion rates and more massive heavy element cores (the effects are linked) result in warm starts, with luminosities matching hot start planets even at young, $10^{6}$--$10^{7}$ year, ages.  \cite{berardo17} studied the structure and role of post-shock (i.e.\ closer to the planet) radiative envelopes.   They found that cold starts could only be produced if the accretion rate was low and the shock temperature was quite cold $\lesssim 500$K, i.e.\ closer to disk temperatures than what is needed to radiate $L_{\rm acc}$.  However, the shock temperature was an input parameter in these models.

To determine the shock temperature and accreted entropy, \cite{marleau19} performed 1D radiation-hydrodynamic simulations of a spherical accretion shock, with realistic opacities to treat dust sublimation fronts.  While they found high radiation efficiencies $> 90\%$, the remainder of the
total incoming energy favors hot starts: high shock temperatures (enough to radiate $L_{\rm acc}$) and post-shock entropy $\sim 13-20 ~k_B/{\rm baryon}$, larger at higher accretion rates and planet masses, without requiring massive cores. \citet{chen22} added more realistic equations of state, and found that H$_2$ dissociation occurs at higher accretion rates, causing a reduction in radiation efficiencies and hotter starts.

While these 1D results support warm or hot starts by core accretion, they assume  spherical accretion onto a $\sim 2 R_{\rm jup}$ planetary surface.  Freely falling material that carries the disk's angular momentum $\sim \Omega R_{\rm H}^2$ should circularize around $\sim R_{\rm H}/3 \gg R_{\rm jup}$ (by factors $> 1000$ for $M_{\rm pl} \gtrsim  M_{\rm Jup}$ at $a \gtrsim 10$ AU).  Thus accretion should occur though CPDs.  Surface accretion shocks could occur if the planetary magnetic field truncates the disk and CPDs are undergoing magnetospheric accretion \citep{lovelace11}. The resulting hot spots could produce optical/UV bump in CPD's spectrum energy distribution \citep{Zhu2015}, which could potentially be constrained observationally. In a model of CPD accretion onto a planetary boundary layer, \cite{owen16} found that hot starts are possible if the CPD is reasonably warm and thick.

While CPDs are being actively studied, current 3D simulations cannot resolve the planetary surface very well (especially when gravitational smoothing lengths are accounted for).\footnote{The isothermal shearing box study of \citealp{Bethune2019}  impressively resolves a rigid planet core, with no smoothing,  enabled by a low density, large radius core, i.e.\ $R_c = H/16 = 67 R_{\rm Jup}$ when the $M_c = 4 M_{\rm th} \simeq  0.5 M_{\rm Jup}$ case is applied to a thin $H/r = 0.05$ AU disk at 10 AU.}  They also tend to lack non-ideal MHD, but see \cite{gressel13}.

With some approximations,  2D simulations can account for rotation and resolve accretion flows and shocks onto CPD and planetary surfaces \citep{takasao21, marleau23}.   However, these studies focus on H$\alpha$ emission, not the evolution of planetary entropy.    More detailed modeling is warranted.  

Observationally, the internal entropy affects planetary luminosity and the gas envelope size,  though the fraction and radial distribution of metals also affects planetary size \citep{baraffe08}.  The internal adiabat  can also affect the chemical species present in the atmosphere, due to mixing that reaches cloud condensation curves \citep{fortney20}.  Observable differences between planets with different initial entropies decrease with age \citep{marley07}, making the characterization of younger planets important for determining initial entropies.

\section{Concluding Remarks}

Some of the main issues concerning the formation of giant planets have been reviewed. The focus has been on the prevailing core accretion hypothesis\index{core accretion hypothesis}, with an explanation of why it is favored. Planets that form by core accretion are expected to have (or have had) significant heavy element cores.  In the Solar System, the  bulk metal fraction of Jupiter is still only roughly constrained to $Z_{\rm pl} \approx 2.5\text{--}14\%$, while $Z_{\rm pl} \simeq 20\pm 1 \%$ for Saturn and $Z_{\rm pl} \approx 75\text{--}90 \%$ for Uranus and Neptune \citep{guillot23}.   These significant fractions, plus the overall trend of decreasing planet metallicity\index{metallicity!of planets} with mass (more gas accretion), are expected for core accretion.

In Jupiter and Saturn, there is evidence that most metals are spread out in a ``fuzzy" core extending to $\sim 0.5 R_{\rm pl}$ \citep{wahl17, mankovich21, nettelmann21}.  However, Jupiter also seems consistent with a uniform $Z_{\rm env} \sim 2\%$ atmosphere and envelope plus a small compact core \citep{nettelmann24}.  Theoretical models have been developed to explain how convection dilutes a compact core into a fuzzy one, especially for high initial entropy and lower planet mass \citep{knierim24}.  

For exoplanets, the bulk densities of giant planets (from transit and radial velocity data) combined with evolution models, suggest a mean exoplanet metallicity $Z_{\rm pl} \sim 0.2 (M_{\rm pl}/M_{\rm jup})^{-0.4}$, with significant scatter \citep{thorngren16}.  This result agrees with the Solar System $Z_{\rm pl}$ vs.\ mass trend, and further suggests that Jupiter is metal poor compared to the average extrasolar Jupiter.  

Exoplanets' atmospheric spectra 
can be fit to an atmospheric metallicity \citep{welbanks19}, which provides a lower bound on the bulk metallicity. 
Saturn mass exoplanets show high \citep{feinstein23}  or very high \citep{bean23} atmospheric metallicities. Jupiter mass planets show both sub- and super-solar  metallicities \citep{line21, fu24}.
The ensemble of exoplanet metallicities adds further support to the core accretion hypothesis. 

Several areas for progress have been identified. Protoplanetary disk observers and theorists are working to better understand the origin of observed disk structures. Models of coupled pebble and planetesimal accretion have the potential to provide new insights, particularly as they more self-consistently account for planetesimal fragmentation, growing gas envelopes, and planet-disk interactions. Models of gas accretion will continue to benefit from the development of 3D radiation hydrodynamic simulations, which incorporate realistic thermodynamics, evolving dust distributions, and improved resolution. While computationally intensive 3D simulations are a key approach, progress can also be made through simplified 2D, 1D, and (semi-)analytic models, which enable higher resolution, more complex microphysics, and broader parameter studies to explore fundamental scalings. The greatest progress is often achieved when insights from observations and theoretical models -- with varying degrees of computational complexity -- flow in multiple directions.

Several important issues not covered in this review are addressed in other reviews.  The migration of forming giant planets occurs on similar timescales to the formation processes described here \citep{nelson18}.  The longer term dynamical evolution of planetary systems, including the particular dynamical evolution of the giant planets in the Solar System, is also highly relevant \citep{morbidelli18}.  Populations synthesis models  combine  prescriptions and/or detailed modeling of many planet formation processes to produce synthetic populations of planets, to then compare to observed populations \citep{mordasini18}.  These issues, and more, arise when trying to explain the origin and evolution of hot Jupiter exoplanets \citep{fortney21}.

While the disks around free-floating\index{free-floating planets} and wide-orbit ($> 100$ AU) planetary mass ($\lesssim 10 ~M_{\rm jup}$) companions were discussed, the origin of these objects was not considered.
They could be ejected from formation in a disk or be the low mass tail of isolated star formation, i.e.\ collapse in a dense molecular cloud filament or core.  Observationally, the stellar and substellar IMF (initial mass function), fit as $dN/dM \propto M^{-\alpha}$, declines to low masses (i.e.\ has $\alpha < 1)$.  JWST observations suggest that $\alpha \sim 0.3$ for $M \gtrsim 12 M_{\rm jup}$,  breaking to a steeper $\alpha \sim -1$ at lower masses, with uncertainty due to low number counts \citep{de-furio24}.   This low mass cutoff in star formation is partly ``opacity limited" by the ability of gas to cool \citep{hennebelle24}.  This serves as a reminder that, occasionally, planetary-mass objects with $\lesssim 5 M_{\rm jup}$ form in a manner similar to isolated stars, a process distinct from both core accretion and disk fragmentation.

\section{Cross-References}
\begin{itemize}
\item{Planet formation theory: an overview \citep{armitage24}}
\item{Planet Occurrence from Doppler and Transit Surveys \citep{winn24}}
\item Instabilities and Flow Structures in Protoplanetary Disks: Setting the Stage for Planetesimal Formation \citep{klahr18HE}
\item{Pebble Accretion \citep{ormel24}}
\item Planetary Migration in Protoplanetary Disks \citep{nelson18}
\item Planetary Population Synthesis \citep{mordasini18}

\end{itemize}


\bibliographystyle{spbasicHBexo}  
\bibliography{giantplanet} 

\printindex

\end{document}